\documentstyle[prl,aps,multicol,epsf]{revtex}

\begin{document}

\draft

\title{Topographic Mapping of the Quantum Hall Liquid using a
Few-Electron Bubble}

\author{G. Finkelstein, P.I. Glicofridis, R.C. Ashoori}

\address{Department of Physics and Center for Materials Science and
Engineering, Massachusetts Institute of Technology, Cambridge, MA
02139}

\author{M. Shayegan}

\address{Department of Electrical Engineering, Princeton University,
Princeton, NJ 08544}

\maketitle

\begin{abstract}

A scanning probe technique was used to obtain a high-resolution map of
the random electrostatic potential inside the quantum Hall liquid. A
sharp metal tip, scanned above a semiconductor surface, sensed charges
in an embedded two-dimensional electron gas. Under quantum Hall effect
conditions, applying a positive voltage to the tip locally enhanced
the 2D electron density and created a ``bubble'' of electrons in an
otherwise unoccupied Landau level. As the tip scanned along the sample
surface, the bubble followed underneath. The tip sensed the motions of
single electrons entering or leaving the bubble in response to changes
in the local 2D electrostatic potential.

\end{abstract}


\begin{multicols}{2}


Since the discovery of the integer quantum Hall effect (QHE) in a
two-dimensional electron gas (2DEG), electron localization has been
proposed to play a key role in the phenomenon\cite{book}. In GaAs
heterostructures, the random potential responsible for localization
derives mainly from randomly situated ionized donors, located between
the submerged 2DEG and the sample surface. Although this picture
remains widely used, it has proven difficult to quantify the random
potential experimentally. Conductivity measurements, the primary tool
for studying the QHE, provide only average information about the
disorder. Recently developed scanning techniques hold promise for
spatially resolving the behavior of semiconductor structures on
nanometer scales\cite{Westervelt,STM,McCormik,Amir,strip}. In
particular, low compressibility strips\cite{CSG} with Landau level
filling factor $\nu$ close to an integer have been imaged directly in
the integer Quantum Hall regime\cite{Amir,strip}. 

We present results from a new technique that images microscopic
details of the disorder potential through detection of motions of
single electrons within the system under study. A {\it mobile quantum
dot} is created inside the 2DEG by enhancing the 2DEG density locally
underneath a sharp scanned metal tip biased to attract electrons
(inset Fig. 2).  While scanning the tip across the sample surface we
drag the dot along underneath. Comparing the single-electron charging
pattern of the dot at different locations, we map the electrostatic
potential directly as sensed by the 2DEG electrons. We find that the
form of this potential remains mostly unchanged for different
spin-split and integer Landau level fillings, which suggests that
screening of the random external potential by the 2D electrons does
not vary appreciably in the quantum Hall regime.

Our 2DEG is formed at the GaAs/AlGaAs interface $80 nm$ below the
sample surface. It has an electron density of $\approx 1.5\times
10^{11}/cm^{2}$ and mobility $\approx 1.5\times 10^{6} cm^2 /Vsec$. A
metallic gate electrode, patterned in a form of a grating covers the
sample surface (Fig. 1). We image a region of the surface several
micrometers in size between two fingers of the gate. In the subsurface
charge accumulation (SCA) method\cite{STM}, a sharp scanning probe is
brought close ($\lesssim 20 nm$) to the surface of the sample. Unlike
scanning tunneling microscopy, no tunneling current passes between the
probe and the sample. Instead, we apply a 100 kHz AC excitation of
$\sim 3$ mV RMS to the 2DEG and to the gate. Because of the
capacitance of the 2DEG to the ground and to the scanning tip,
electric charge flows in and out of the 2DEG. We monitor this charging
of the 2DEG by measuring the image charge induced on the scanning
probe by a sensitive cryogenic amplifier.  

In scanning probe measurements of semiconductor and nanoscale
structures the measuring tip may, itself, strongly alter the local
electron density and thus produce artifacts in images. Some
researchers have taken advantage of this effect by using a scanning
tip to alter the current flow in nanoscale systems\cite{Westervelt}.
In previous work using SCA microscopy, we compensated the work
function difference between the scanning tip and the sample to avoid
the tip perturbing the sample electrostatically\cite{STM}. The
resulting SCA images revealed features attributed to variations of the
2DEG compressibility. In the present work, we instead purposefully
apply a voltage on the scanning tip to change the electron density
underneath. For magnetic fields such that the bulk Landau level
filling factor is slightly less than integer, we thereby create a
``bubble'' of electrons in the next Landau level. The electrons in the
bubble are separated from the surrounding 2DEG by an incompressible
strip of integer Landau level filling. The imaged features do not
display the compressibility variations inherent to the 2DEG. Instead,
Coulomb blockade in the bubble determines the observed signal, and the
images display a set of equipotential contours that serve as a
topographic map of the random electrostatic potential in the 2DEG. One
may "read" directly the electrostatic potential by noting the
positions of the contours.

We find the voltage that compensates the electric field between the
tip and the sample by measuring the Kelvin signal (see Fig.
1)\cite{Kelvin}, which is proportional to the {\it electrostatic}
potential difference between the tip and the sample. With no external
bias applied between them, the electrostatic potential difference
equals to the work function difference between the two. We null it by
applying an opposing bias to eliminate any electric field between the
tip and the sample. Henceforth, we designate the tip-sample voltage as
measured in deviation from this nulling voltage.

In the SCA image of a $2.5 \times 2.5 \mu m$ region at magnetic fields
of $5.8$ T, corresponding to the bulk Landau level filling factor $\nu
=1$ (Fig. 2a), regions of high and low signal are presented as bright
and dark colors, respectively. The voltage between the tip and the
sample is $+1$ V. This corresponds to accumulation of the electrons
underneath the tip. We observe a strikingly complex pattern of closed
contours, reminiscent of a topographic map. The contours evolve
rapidly with magnetic field, staying nested around fixed centers.

The size of the smallest feature within 2DEG resolvable by the SCA
microscopy is limited by the depth of the 2DEG, $d=80$nm. However, we
measure separations between contours lines as small as $50 nm$.
Therefore, we conclude that the contours do not represent charging
patterns inherent to the 2DEG. Rather, the features reflect variations
in the 2DEG charging induced by changes of the tip position. The
periodic appearance of the contours hints that they originate from
Coulomb blockade\cite{QDs}. Indeed, when we fix the tip position and
vary the tip-sample voltage, the signal displays periodic
oscillations. In this situation, the tip acts as a gate controlling
the electron number in a quantum dot.

We explain the observed contours as follows. An electron accumulation
underneath the tip forms a few-electron bubble (inset Fig. 2). As the
bubble is dragged in the 2DEG plane following the tip, it experiences
different local electrostatic potentials. When the potential energy
for electrons is high, electrons are expelled from the bubble; when
the potential energy is low, electrons are drawn into the bubble. As
the tip scans across these positions of single electron transfer, the
applied AC excitation causes an electron to move back and forth
between the bubble and the surrounding 2DEG, producing a peak in the
synchronously-detected SCA signal\cite{Ray}. Between the peaks,
Coulomb blockade prohibits transfer of electrons to and from the
bubble, and a minimum of SCA signal is detected. As a result, SCA
images display alternating contours of high and low SCA signal
surrounding minima and maxima of the random electrostatic potential
within the 2DEG.

The contours observed in the experiment do not intersect and have
roughly the same contrast across the image. This confirms that they
originate from a single bubble located underneath the
tip\cite{comment1}. The fact that contours appear only on the high
magnetic field side of the Quantum Hall plateau also supports this
scenario. In Figure 2c, we present four images at different magnetic
fields around $\nu = 1$. For the top three images, the Landau level
filling factor in the bulk of the sample is $\nu < 1$, while
underneath the tip it is $\nu > 1$. The bubble of $\nu > 1$ is
separated from the bulk by the incompressible strip with $\nu=1$,
(inset, Fig. 2). The strip serves as a barrier between the bulk 2DEG
and the bubble and ensures charge quantization\cite{vandervaart}. In
contrast, at the lower field side of the Quantum Hall plateau (lowest
image, Fig. 2c) the filling factor is everywhere larger than 1 and a
mere density enhancement forms, rather than a bubble capable of
trapping single electrons. 

We show the evolution of SCA images at 6 T with different tip-sample
voltages in Fig. 2d. As the voltage decreases (lower images), the
contours shrink and disappear around nulling voltage\cite{future}.
Without electric field between the tip and the sample, the presence of
the tip does not influence the density distribution in the 2DEG, so
that the electron bubble does not form and no contour lines
appear. We confirmed this statement by studying the SCA
images measured at different tip heights above the surface; away from
nulling voltage, the images depend strongly on changing the tip
height, while at nulling voltage they stay virtually unchanged.

In some cases, submicrometer structures similar to those discussed
above persist at any voltage between the tip and the sample, including
the nulling voltage for the Kelvin signal. By measuring the Kelvin
signal as a function of the tip height, we have found that in these
cases the tip does not have a uniform workfunction over its entire
surface. Most probably, this results from GaAs debris partially
covering the scanning tip. As a result, an electric field exists
between the probe and the sample at any tip-sample voltage, even when
the averaged Kelvin signal is zero. We therefore carefully checked the
cleanliness of the tip throughout the experiments reported here.
Possibly, in previous work \cite{STM}, the tip-to-sample electric
field was not completely eliminated and debris on the tip resulted in
a non-uniform workfunction difference between different regions of the
tip and the sample. Therefore, some of the features (particularly arcs
and filaments) observed in Fig. 2 of \cite{STM} may be artifacts
resulting from the tip locally perturbing the sample. 

Imaging of another region of the sample at three different magnetic
fields and tip-sample voltages is shown in Fig. 3. Each contour line
marks a position within the 2DEG plane, where the number of electrons
in the bubble changes by one. As the tip voltage decreases, the energy
of the $N$ electron state drops relative to the $N+1$ electron state,
and the contour line moves. Monitoring the evolution of the images at
a fixed field, we observe that contours shrink around the central
locations as the voltage decreases towards nulling. We conclude that
inside each contour the electron bubble has one more electron than
outside. In particular, the contours surround two local minima of the
potential (as sensed by electrons). Note that different contours
follow the same evolution as we change the tip voltage to move one
contour to a position formerly occupied by another. For example, at
$B=6.5T$ the inner contour at $+1.0V$ and the outer contour at $+0.5V$
appear indistinguishable in size and shape. This confirms our view of
the bubble as an electrostatic potential probe; despite differences in
bubbles created by different tip voltages, the resulting contours
remain practically unchanged.

Our equipotential contours have the same meaning as topographic
contours on a land map. To measure the amplitude of the electrostatic
potential inside the 2DEG, we need to know the energetic separation
between the single-electron contours. This information comes from
measuring the width of the single electron peaks in a SCA trace taken
as a function of the tip voltage at a fixed location. We fit the shape
of the single electron peaks by the derivative of the Fermi
distribution\cite{QDs}. Assuming the that width of the single-electron
peaks is determined by the temperature of $0.35 K$, we find that the
Fermi energy at the bubble is changed by $\sim 2 meV$ per change of
the tip voltage by $1 V$. Also, as we change the tip voltage by $1 V$,
from $+0.5V$ to $+1.5V$, at $6.6T$ the innermost contour expands to
span the entire area of Figure 3. We thus conclude that the range of
the random potential in this region is about $2 meV$.

The contour lines observed near bulk filling factor $\nu =1$ reappear
around $\nu =2$ (compare Fig. 2a and b). The contours observed at $\nu
=1$ display a stronger contrast, although the expected value of the
exchange-enhanced spin gap ($\nu =1$) is smaller that the cyclotron
gap ($\nu =2$). Tunneling of an electron between the interior and
exterior of the bubble should be additionally suppressed close to $\nu
=1$ by the opposite spin orientation of the transferred electron in
the initial and final states. This suppression of tunneling does not
exist close to $\nu =2$, and as a result the Coulomb blockade may be
less pronounced in the latter case.

The contours clearly encircle the same locations at $\nu =1$ and $\nu
=2$. The close similarity between the images indicates that we observe
a fingerprint of the same random electrostatic potential at both
filling factors. The random potential fluctuations extend laterally by
typically $\sim 0.5 - 1 \mu m$. This scale significantly exceeds $50
nm$, the width of the spacer layer that separates 2DEG from donors.
Random potential fluctuations due to remote ionized donors should have
a characteristic lateral scale of about the spacer width
\cite{Efros1}. Most probably, screening by the residual electrons left
in the donor layer smoothens the potential in our sample. Note, that
the size of the scanned bubble, $\sim 100 nm$ (see below), in
principle does not prevent observation of smaller scale potential
fluctuations. Indeed, at select locations the single electron contours
display very small radii of curvature, signaling steep potential bumps
or dips (Fig. 3).

Figure 3 also traces the evolution of contours with magnetic field.
Interestingly, the changes induced by variations of magnetic field and
the tip-sample voltage are quite similar. In fact, we can compensate
for the change of a contour's size induced by magnetic field by
properly tuning the tip-sample voltage (Fig. 3, images along the
dashed diagonal line). The size of contours remains constant at
$\nu=1$ along lines of ${dV_{tip}\over dB} \approx 2$ V/T. At $\nu=2$
the constant shapes of contours are instead preserved at
${dV_{tip}\over dB} \approx 4$ V/T. We can readily explain this
observation by recalling that the bubble is formed on top of an
integer number of completely filled (spin-split) Landau levels. When
magnetic field increases, the degeneracy of these levels grows, and
they can accommodate more electrons. To supply these electrons, we
need to apply a larger voltage between the tip and the sample. We
expect the voltage required for compensation to be roughly
proportional to the number of filled Landau levels, in agreement with
the experiment.

We determine the size of the bubble by measuring the periodicity of
the signal with magnetic field. Adding one flux quantum per area of
the bubble adds one electronic state to each filled (spin-split)
Landau level, removing roughly an electron from the upper, partially
filled Landau level that constitutes the bubble. We observe that as
the magnetic field is changed by $\approx 0.2 $T, the contours shift
by one complete period. In Figure 3, the inner contour at $B=6.4$T
coincides with the outer contour at $B=6.6 $T. For the case of Figure
2 the magnetic field period has a similar value of $\approx 0.15$ T,
implying an area of $0.02 \mu m^2$. This compares favorably with the
square of the 2DEG depth. Because the magnetic field period does not
depend on location within the image, it appears that the size of the
bubble is not strongly affected by the random potential.

In conclusion, we have formed a mobile quantum dot inside the 2DEG by
locally accumulating electron density underneath the scanning probe.
By comparing the Coulomb blockade patterns at different locations we
map the potential inside the 2DEG as sensed by the electrons. We find
that the 2D electron screening of the random potential induced by
external impurities changes little between different quantum Hall
plateaus and within each plateau. With demonstrated single-electron
sensitivity, our subsurface charge accumulation may allow
understanding of a wide variety of submerged electronic structures on
the nanometer scale.

\figure Figure 1: Sample and measurement schematics. Upper panel
describes the charge accumulation (capacitance) method with an AC
excitation applied to the sample and the gate. Capacitive coupling
between the tip and the 2DEG induces an AC signal on the tip. Lower
panel: Capacitance image of an $11 \times 11 \mu ^{2}$ region,
including two ``fingers'' of the gate. The gate resides directly on
the surface of semiconductor, while the 2DEG is buried underneath the
surface. Therefore, the gate produces a larger capacitance signal. In
the Kelvin probe method we vibrate the sample in the vertical
direction with a frequency of $2 kHz$ and an amplitude of $\sim 10
nm$. We measure the AC charge induced on the scanning probe. It is
proportional to $V\frac{dC}{dz}$, where $C$ is the probe-sample
capacitance, $z$ is the tip height and $V$ is the {\it electrostatic}
potential difference between the tip and the sample.

\figure Figure 2a: $2.5 \times 2.5 \mu m$ SCA image at $5.8$ T ($\nu
\lesssim 1$) and $V_{tip}= +1$ V. b: Image of the same region at $3.0$
T ($\nu \lesssim 2$). c: Magnetic field evolution of the feature at
the top left corner of Fig. 2a at $V_{tip}= +1$ V. Top to bottom:
$6.5, 6.1, 5.9$ and $5.7$ T. The features disappear when more than one
spin-split Landau level is filled in the bulk ($\nu > 1$). d:
Evolution of the same feature with tip voltage, $6.0$ T. Top to
bottom: $1.0, 0.7, 0.4$ and $0.0$ V. At the nulling voltage (lowest
image) the tip does not influence 2DEG. Inset: schematic of the 2DEG
density profile underneath the tip. Note the ring of constant density
that corresponds to the incompressible strip with an integer Landau
level filling factor (inset).

\figure Figure 3: Series of $2 \times 2 \mu m$ SCA images taken at a
different location from Fig. 2. The changes in contour size introduced
by the tip voltage and magnetic field compensate each other. See
images along the diagonal dashed line. Lower panel: Schematic
explanation of the changes induced in the bubble by magnetic field and
the tip voltage. 

\end{multicols}


\begin{references}
\bibitem{comment1} Single electron features may also be observed in
the SCA imaging if there exist localized ``puddles'' of electrons
formed within the quantum Hall liquid. In this case, a biased tip
scanning across a puddle would act as a gate, changing the number of
electrons in it. The SCA images would then exhibit concentric rings of
high and low SCA signal. Since each puddle would have its own distinct
series of electron additions, the set of rings from different puddles
would superimpose. Experimentally, the bright and dark bands form a
single network of contour lines that never intersect. Therefore, we
conclude, that there exists only one few-electron bubble formed under
the tip.
\bibitem{future} Contour lines reappear when we apply negative
tip-sample voltages depleting the 2DEG at magnetic fields $B<6$ T,
when more than one Landau level is filled in the bulk of the sample.
Here we attribute them to a Coulomb blockade in a bubble formed by
quasi-holes in the filled Landau level.
\end{references}
\end{document}